\documentclass[aps,pra,twocolumn,amsmath,amssymb,superscriptaddress]{revtex4-1}
\usepackage{graphicx}
\usepackage{xcolor}
\usepackage{threeparttable}
\usepackage[colorlinks=true,
					linkcolor=blue,
					anchorcolor=blue,
					citecolor=blue,
					urlcolor=magenta]{hyperref}
\usepackage{hyperref}

\begin{document}

\title{Leveraging junk information to enhance the quantum error mitigation}

\author{Ruixia Wang}
\affiliation{Beijing Academy of Quantum Information Sciences, Beijing 100193, China}

\author{Xiaosi Xu}
\affiliation{Graduate School of China Academy of Engineering Physics, Beijing 100193, China}

\author{Fei Yan}
\affiliation{Beijing Academy of Quantum Information Sciences, Beijing 100193, China}

\author{Xiaoxiao Xiao}
\affiliation{Key Laboratory of Theoretical and Computational Photochemistry, Ministry of Education, College of Chemistry,
Beijing Normal University, Beijing 100875, China}

\author{Ying Li}
\email{yli@gscaep.ac.cn}
\affiliation{Graduate School of China Academy of Engineering Physics, Beijing 100193, China}

\author{Xiaoxia Cai}
\email{caixx@baqis.ac.cn}
\affiliation{Beijing Academy of Quantum Information Sciences, Beijing 100193, China}

\author{Haifeng Yu}
\affiliation{Beijing Academy of Quantum Information Sciences, Beijing 100193, China}
\affiliation{Hefei National Laboratory, Hefei 230088, China}

\date{February 2024}

\begin{abstract}
Noise in quantum information processing poses a significant obstacle to achieving precise results. Quantum error mitigation techniques are crucial for improving the accuracy of experimental expectation values in this process.  
In the experiments, it is commonly observed that some measured events violate certain principles, such as symmetry constraints. These events can be considered junk information and should be discarded in a post-selection process. 
In this work, we introduce a quantum error mitigation method named Self-Trained Quantum Noise Filter (SQNF), which leverages the junk information to differentiate errors from the experimental population distributions, thereby aiming to approximate the error-free distribution. 
Our numerical results demonstrate that the proposed method can significantly reduce the infidelity of population distributions compared to the traditional post-selection method. 
Notably, the infidelity reduction is achieved without additional experimental resource consumption. 
Our method is scalable and applicable to multi-qubit computing systems. 
\end{abstract}

\maketitle

Quantum computing has gained significant attention in recent years due to its potential to revolutionize various fields, including cryptography \cite{Shor_1999}, machine learning \cite{Cerezo_2022,David_2024,Biamonte_2017}, and chemistry \cite{McArdle_2020,Cao_2019,Abrams_1999,Paesani_2017,Peruzzo_2014,shang_2024}. 
Many of these applications use deep circuits, such as the Quantum Fourier Transformation (QFT) \cite{Nielsen_2010}. For this reason, they are generally considered feasible for fault-tolerant quantum computing that utilizes the quantum error correction (QEC) \cite{Georgescu_2020,Sivak_2023}. Since the technologies of quantum error correction are yet under development, we are still in the Noisy Intermediate-Scale Quantum (NISQ) era \cite{Preskill_2018}, where the noise from devices and the environment can significantly impact experimental results. However, valuable computations can still be implemented on NISQ devices utilizing an alternative method such as quantum error mitigation (QEM) \cite{Cai_2023, Endo_2021}. Unlike QEC, QEM reduces the bias in experimental data obtained from noisy quantum circuits by various methods, such as post-processing data and increasing the noise in circuits.

Several QEM methods have been proposed in previous works. In the zero-noise extrapolation (ZNE) method \cite{Li_2017,Temme_2017}, one amplifies the circuit noise and estimates the result in the zero-noise limit through polynomial or exponential function fitting. In the probabilistic error cancellation (PEC) \cite{Temme_2017,Endo_2018}, the circuit is randomised to compensate for the bias caused by noise. Both two methods can approach unbiased results as long as an accurate noise model is provided. In experiments, the noise model is usually obtained via a set of benchmarking circuits \cite{Kandala_2019,Kim_2023,Chao_2019,Zhang_2020,van_2023}, which could be a significant time cost. It is similar in learning-based methods \cite{Strikis_2021,Czarnik_2021}, in which extra circuits are executed to find optimal parameters that are eventually used in QEM. Some other methods, such as symmetry-based post selection \cite{McArdle_2019,HF_2020}, subspace expansion method \cite{McClean_2017,
McClean_2020,Takeshita_2020} and virtual distillation \cite{Koczor_2021,Huggins_2021}, still work without the benchmarking stage. However, they can only reduce certain types of errors. For example, the symmetry-based method is widely applicable in quantum computational chemistry and many-body physics calculations \cite{Tilly_2022}. When dealing with these systems, quantum circuits always exhibit intrinsic symmetries due to the physical nature of the object, such as the conservation of particle number, parity or spin. By post-processing data, one can reduce errors that break the symmetries.

This work introduces a QEM method named the Self-Trained Quantum Noise Filter (SQNF). In this method, we utilise the junk data discarded in the post-selection to benchmark the noise model. Compared with the post-selection method, our method can reduce all types of errors, including those breaking and preserving the symmetries. Meanwhile, extra benchmarking circuits are not required, which greatly simplifies the experimental implementation of QEM.

\begin{figure}[htbp]
\centering

\includegraphics[scale=1]{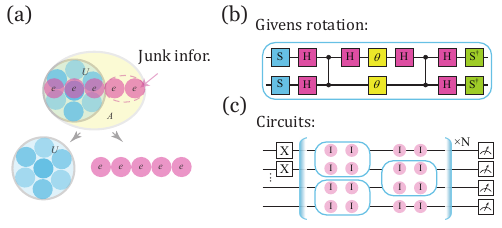}

\caption{\label{f1} (a) The diagram illustrates the conceptual framework of our proposal. Set A encompasses the complete information of a system. The subset U contains the useful information, which is inevitably mixed with errors denoted by e. Correlations exist between the errors inside and outside of the subset U. By acquiring information about the error e outside the subset U, labeled as `Junk infor.', we can deduce the error e within U. Consequently, we are able to separate the useful information and errors within subset U, obtaining the desired information. 
(b) Our circuit design employs elementary units known as `Givens rotations,' which maintain the excitation number in the quantum circuits. Each Givens rotation consists of two rotation operations along the $\sigma_y$ axis, characterized by a random parameter $\theta$, depicted as yellow squares in the figure. 
(c) The circuits are constructed using Givens rotations, represented by blue squares. During simulations, we introduce noise channels on the identity gates, uniformly inserted into the circuits as indicated by the pink circles.
}
\end{figure}

In the quantum computation and simulation process, if there is predicted information available, such as illustrated in Fig. \ref{f1}(a), where we identify that the useful information resides within a subset U (where U is a subset of the complete set A), and there exist correlations for the errors, denoted by e, both within and outside the subset U, we can effectively segregate the errors and useful information within subset U by analyzing the correlations between the errors inside and outside of U. 
The errors outside of subset U can indeed be considered as junk information since they appear to be irrelevant and are typically discarded directly using conventional methods.
However, in our proposal, we aim to leverage this junk information to facilitate quantum error mitigation techniques.

In quantum computing processes, noise is an inherent factor within quantum circuits. For extensive circuits comprising numerous qubits and quantum gates, as the circuit length expands, the noise model can be approximated effectively by the depolarizing noise channel \cite{urbanek2021mitigating,vovrosh2021simple,he2020zero}. If we construct the noise model using global depolarizing noise channel, the final density matrix after running $n$ layers of the quantum circuit can be represented as:

\begin{equation}
\rho_f = \frac{P_nI}{d}+(1-P_n)\rho_{ideal},
\end{equation}
where $P_n = 1-(1-p)^n$ represents the equivalent global depolarizing rate after $n$ layers of the circuit. $p$ is the global depolarizing rate for each circuit layer. 
$d$ is dimension of the density matrix. $I$ is the identity matrix with the dimensions $d\times d$. 
$\rho_{ideal}$ is the ideal density matrix without noise in the quantum circuit, and $\rho_f$ is the final density matrix after running $n$ layers of the quantum circuit with the depolarizing noise channel. 
The useful information is contained within $\rho_{ideal}$, and if $\rho_{ideal}$ can be written as a block matrix, such as

\begin{eqnarray}
\rho_{ideal} = \left[\begin{array}{cc}
 \rho_{u} & 0_m \\
  0_m & 0_m
\end{array}\right],
\label{eqM}
\end{eqnarray}
where $\rho_u$ is a $d_u\times d_u$ matrix with $d_u< d$ and contains all the non-zero information in $\rho_{ideal}$, here we define the subspace of $\rho_u$ as the useful subspace, and indicated by $S_u$. Correspondingly, the space containing full information of $\rho_f$ is represented by $S_a$. $0_m$ is the zero matrix. Then we have
\begin{eqnarray}
\rho_{f} = \frac{P_n}{d}\left[\begin{array}{cc}
 I_u & 0_m \\
  0_m & I_m
\end{array}\right]+(1-P_n)\left[\begin{array}{cc}
 \rho_{u} & 0_m \\
  0_m & 0_m
\end{array}\right].
\label{eqM}
\end{eqnarray}

We can define that $\tilde{\rho}_{u}=cI_u+(1-P_n)\rho_u$, which represents the final density matrix for the useful subspace with errors. And $c=\frac{P_n}{d}$ is a constant value depending on the strength of the depolarization noise in the circuits. 
In the realistic experiments, $\tilde{\rho}_{u}$ can be obtained by the measurements, if the constant value of $c$ can be obtained, the ideal density matrix of $\rho_u$ can be extracted as 
\begin{equation}
\rho_u=\frac{\tilde{\rho}_u-cI_u}{1-P_n}.
\end{equation}
In the final matrix $\rho_f$, there is another block matrix at the bottom right, which can be defined as $\rho_J=cI_m$ and the subspace related to this matrix is indicated as $S_J$. It should be noted that $S_u \oplus S_J = S_a $. Typically the results in $\rho_J$ is discarded directly, because it contains no useful information. However, we have discovered that, if the noise model can be approximated as the global depolarizing noise channel, we can extract the constant value $c$ from the junk data in $S_J$. Then we can separate the useful information and the noise induced errors in the matrix $\tilde{\rho}_u$. In this way, the depolarizing noise induced errors can be filtered out without running extra circuits or any extra measurements, so that, we named it as the Self-Trained Quantum Noise Filter (SQNF).

In this work, to validate the efficacy of our SQNF theory, we utilise Givens rotations with random parameters $\theta$ as the elementary units to generate the quantum circuits, as depicted in Fig. \ref{f1}(b). The circuits generated by Givens rotations maintain the excitation number, implying that if the initial state has an excitation number of $n_e$, then ideally, we should only detect populations in states with a total excitation number of $n_e$. 
Here we define the subspace maintaining the conservation of the excitation number as $S_c$ and the complementary set of $S_c$ is $S_{nc}$. In this scenario, the $S_c$ subspace is larger than or equivalent to the useful subspace $S_u$, we can denote it as $S_u\subseteq S_c$.
The circuits, consisting of $n_q$ qubits and $N$ circuit layers is shown in Fig. \ref{f1}(c). Each layer comprises $n_q-1$ Givens rotations and the related random parameters. 
The local depolarizing noise channel is applied uniformly to the identity gates implemented on the circuit. The simulation results, including those with amplitude damping and phase damping noise on one- and two-qubit gates, are depicted in the last figure and also discussed in Appendix \ref{app:sec1}.

\begin{figure}[thbp]
\centering

\includegraphics[scale=1]{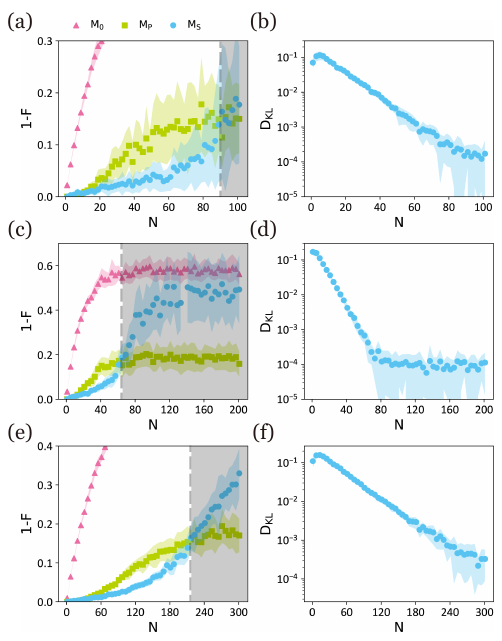}

\caption{\label{f2} The infidelity $1-F$ and the corresponding KL divergence over circuit length $N$. $F$ is the fidelity for the population distribution of the final states. $D_{KL}$ is for the states in the subspace $S_{nc}$. The depolarizing noise rates $\gamma$ on every identity gate, qubit number $n_q$ and excitation number $n_e$ are $\gamma = 0.005$, $n_q = 4$, $n_e = 2$ for (a) and (b); $\gamma = 0.005$, $n_q = 6$, $n_e = 3$ for (c) and (d); $\gamma = 0.001$, $n_q = 8$, $n_e = 4$ for (e) and (f). $\rm M_0$ represents the results without any error correction. $\rm M_P$ represents the method which discards the populations of the states within subspace $S_J$ directly. $\rm M_S$ indicates the method SQNF, which filtered out the depolarizing noise with the help of the junk data. The data shown with markers represent the average values of 21 random circuits. The areas filled with colors show the error bars, which are obtained by calculating the standard deviations with 21 random circuits. In the gray shadows, the infidelities with the three methods become flat, which reveals that, the final states in this area approach the depolarizing error induced maximally mixed states.
}
\end{figure}

The numerical simulation results with the qubit numbers $n_q=4,6,8$ are presented in Fig. \ref{f2}. The fidelity of the measured results is calculated using the following expression $F=\sum_{i=1}^{d}\sqrt{p_i\cdot p_i^{ideal}}$, where $p_i^{ideal}$ is the ideal population, $p_i\in P_{M_0}, P_{M_P}$ or $P_{M_S}$ is the population for the final states with different post processing methods $\rm M_j$, $\rm j=0,P,S$. For $\rm M_0$, the populations in $P_{M_0}$ are raw data without any correction.
For $\rm M_P$, we discard the populations outside of the useful subspace directly. Subsequently, we normalize the populations within the useful subspace to obtain the set $P_{M_P}$. This method is commonly employed in post-selection techniques.
For $\rm M_S$, which is also named SQNF, we leverage the populations discarded by $\rm M_P$, and employ a filtering process to remove the depolarizing noise from the useful data.
With this method, we first calculate the constant value $c$ by averaging the populations outside of the useful subspace. Then we set the populations outside of the useful subspace to zero. Next, we subtract the constant value $c$ from the populations within the useful subspace. If any negative values appear during this step, we also set them to zero. Finally, we normalise the populations within the useful subspace to obtain the set $P_{M_S}$. This process for $P_{M_S}$ can be expressed as
\begin{equation}
\label{eq:mj}
\begin{split}
c &= \frac{\sum_{i=1}^{N_J}p^J_i}{N_J},\\
p_i^J &\leftarrow 0,\, i = 1,2,...,N_J,\\
p_i^u &\leftarrow \max(0,p_i^u-c), \, i = 1,2,...,N_u,\\
p_{sum}^u &=\sum_{i=1}^{N_u}p^u_i,\\
p_i^u &\leftarrow \frac{p_i^u}{p_{sum}^u}, \, i = 1,2,...,N_u,
\end{split}
\end{equation}
where $N_J$ and $N_u$ are the numbers of the states in $S_J$ and $S_u$, and $N_J+N_u=d$. $p_i^x$ is the population for the ith state in the subspace $S_x$, $x=J$ or $u$.

To benchmark the degree of the depolarization for the final states, we use the KL divergence which is defined by 
\begin{equation}
D_{KL} = \sum_{i=1}^{N_{nc}^s}p_i^{even} \log \frac{p_i^{even}}{p_i^{ncs}}.
\end{equation}
For the small size of the space $S_a$ and large circuit noise, we can set $N_{nc}^s=N_{nc}$, and $p_i^{ncs}$ can be written as $\frac{p_i^{nc}}{\sum_{j=1}^{N_{nc}}p_j^{nc}}$ with $p_i^{nc}\neq 0$, where $N_{nc}$ is the number of states in $S_{nc}$ and $p_i^{nc}$ is the population for the states in $S_{nc}$. $p_i^{even} = \frac{1}{N_{nc}}$ is the probability for the normalized even distribution. However, when the space $S_a$ becomes larger or the noise is relatively small, with finite number of the sampling for the measurements, there may appear $p_i^{nc}=0$ for certain state populations. To address this problem and ensure scalability for our proposal, we can divided the states in $S_{nc}$ into several groups, denoted as $G_{nc} = \{G_1, G_2,...,G_{N_{nc}^s}\}$. Each $G_j\in G_{nc}$ contains several quantum states in $S_{nc}$, and $N_{nc}^s$ is the total number of elements in $G_{nc}$. And $p_i^{ncs}$ represents the normalized average probability for the populations in every group $G_i\in G_{nc}$. This approach allows us to calculate KL divergence for the space $S_{nc}$ efficiently, even with large quantum systems.

\begin{figure}[tbp]
\centering

\includegraphics[scale=1]{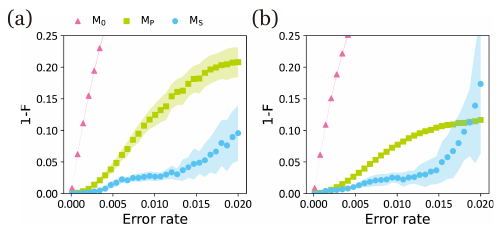}

\caption{\label{f3} The infidelity of the population distribution over the depolarizing error rate. The errors are added on the identity gates. The effectiveness of the error mitigation method is affected by the population distribution of the states in $S_u$. The qubit number and excitation number are $n_q = 4$ and $n_e = 2$, respectively. the circuit depth is set as $N=20$. The KL divergences for the population distribution of the states in $S_u$ are (a) $D_{KL}^{u}=1.85$. (b) $D_{KL}^{u}=0.88$.}
\end{figure}

As shown in Fig. \ref{f2}(a), the infidelities with $\rm M_P$ and $\rm M_S$ are far less than $\rm M_0$ across all length ranges. Comparing $\rm M_P$ and $\rm M_S$, at the initial values of the horizontal axis $N$, the infidelities with $\rm M_S$ may not be lower than $\rm M_P$, that is because the noise model applied on the circuits is a local but not a global depolarizing noise channel, and it requires several layers to enhance the degree of the global depolarization for the circuits. 
After the initial stage, as the circuit depth increases, the average infidelity with $\rm M_S$ becomes significantly smaller than $\rm M_P$. 
Around $N=90$, there is an intersection point for $\rm M_S$ and $\rm M_P$, which is indicated by a dashed line. 
After this point, the tendencies of $\rm M_P$ and $\rm M_0$ [not shown in Fig. \ref{f2}(a), but shown in (c)] gradually flattening out,  indicating that the final states of the circuits approaches the maximally mixed states. The corresponding values of the KL divergence also decrease from around $10^{-1}$ in the beginning to below $10^{-3}$ at the intersection point as shown in Fig. \ref{f2}(b). 
The reduction of the KL divergence indicates the evener distribution of the populations for the states in $S_J$. 
Because the probabilities for the bit flip errors with one or more qubits are different, for example, the probabilities for the errors of $0011\rightarrow 0111$ and $0011\rightarrow 1111$ are different, when $N$ is small, the probabilities of the states in $S_{nc}$ are significantly different. 
As the circuit depth increases, the noise channel approaches the global depolarizing noise, resulting in the probability distribution approaching an even distribution. That is the reason for the reduction of KL divergence.
To enhance scalability, we  partition the states in $S_{nc}$ into three sets based on the excitation numbers: $G_1 = \{0,4\}$, $G_2 = \{1\}$, $G_3 = \{3\}$. 
The population $P_i^{ncs}$ for each group $G_i$ is the average result over all the elements in $G_i$ and then normalized for all $P_i^{ncs}$. 
Thus, in this case, we have $N_{nc}^s=3$. The KL divergence is calculated from the evenly distributed populations $\{\frac{1}{3},\frac{1}{3},\frac{1}{3}\}$ and $\{p_1^{ncs},p_2^{ncs},p_3^{ncs}\}$ with $\sum_i p_i^{ncs}=1$.

In Fig. \ref{f2}(c) and (d), we present the simulation results with 6 qubits from $N=1$ to $N=200$. 
After the intersection point between $\rm M_S$ and $\rm M_P$, the tends of the infidelities with the three methods become flat and correspond to three different values. These three distinct stable values may arise from three different data processing methods. 
The ideal population distribution is an ensemble of random probability distributions constrained in the subspace $S_c$ resulting from the random circuits generated from Givens rotations. We can represent it as $P_c^{ens}$. 
For $\rm M_0$, we do not perform any processing with the raw data. When the circuit depth increases, the final population approaches an even distribution across the entire space $S_a$. We represent it as $P_a^{even}$.
For $\rm M_P$, we set the populations with the states outside the useful subspace as zero, and normalize the populations for the states within the useful subspace $S_u$. When the circuit becomes deep, the final distribution after post processing approaches an even distribution in the subspace $S_u$. We represent it as $P_u^{even}$.
For $\rm M_S$, we perform the data processing using Eq. \ref{eq:mj}. When the circuit becomes deep, after implementing the data post processing, the number of non-zero final populations in the useful subspace will decrease, the normalized final populations will approach a concentrated distribution in the subspace $S_u$. We represent it as $P_u^{con}$. 
The fidelity for $\rm M_0$, $\rm M_P$ and $\rm M_S$ are the average fidelity between $P_c^{ens}$ and $P_a^{even}$, $P_u^{even}$ and $P_u^{con}$ respectively. That is the reason for the three different stable values.
Fig. \ref{f2}(e) and (f) are the results for 8 qubits with a depolarizing rate of $10^{-3}$. The results demonstrate that, as the qubit number increases, the superiority of $\rm M_S$ is also significant. 
With relatively low error rate, the errors for the final states with larger depth of the circuits can also be mitigated effectively. 
Comparing the figures for the infidelities and KL divergences, we observe that, the KL divergence offers insights into the degree of the depolarization for the final states and it can act as an indicator for the applicability  of $\rm M_S$.

The difference in infidelities between $\rm M_P$ and $\rm M_S$ is also influenced by the ideal population distributions of the states in the subspace $S_u$.
As depicted in Fig. \ref{f3}(a) and (b), the results are for two different random circuits with varying KL divergences for the population distributions of the states in $S_u$, which can be calculated as
\begin{equation}
D_{KL}^{u}=\sum_{i=1}^{N_{u}}p_i^{even} \log \frac{p_i^{even}}{p_i^{ideal}},
\end{equation}
where $p_i^{even}$ is the normalized even population distribution for $N_u$ elements, and $p_i^{ideal}$ is the ideal population distribution for the states in $S_u$.
We can observe that, with a higher value of $D_{KL}^{u}$, the superiority of $\rm M_S$ becomes more significant. This indicates that, when the population distribution of the ideal states is far away from an even distribution, error correction with $\rm M_S$ will become more efficient.

\begin{figure}[tbp]
\centering

\includegraphics[scale=1]{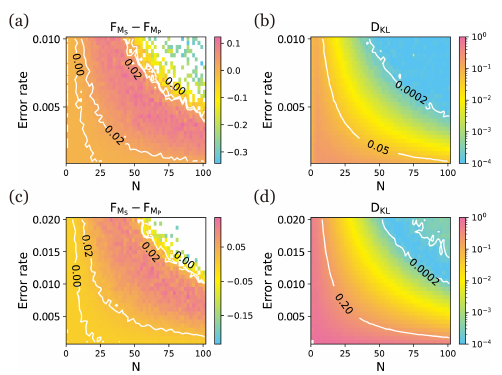}

\caption{\label{f4} The difference of the fidelity $F_{M_S}-F_{M_P}$ and KL divergence $D_{KL}$ is shown for different circuit depth $N$ and error rates, where $F_{M_i}$ represents the fidelity of the population distribution with the method $\rm M_i$, $\rm i=S$ or $\rm P$ with the qubit number $n_q=4$.
For (a) and (b), the error model is the depolarizing noise channel on the identity gates. The error rate is shown on the vertical axis.
For (c) and (d), the error model is the amplitude damping noise and phase damping noise on every single- and two-qubit gates without identity gates. 
The error rate shown on the vertical axis is $\gamma_p$, which is the phase damping noise rate on two-qubit gates. And the amplitude damping rate for two-qubit gates is set as $\frac{\gamma_p}{10}$. The corresponding error rate for single-qubit gates is set with $\frac{1}{5}$ of which for two-qubit gates.
}
\end{figure}

In the above simulation results, we know that, the KL divergence provides information regarding the efficacy of $\rm M_S$. To show the similarity in pattern between the fidelity difference of $F_{M_S}-F_{M_P}$ and the KL divergence $D_{KL}$, we  present the heat map in Fig. \ref{f4}. The upper-right areas in Fig. \ref{f4}(a) and (c) indicate that the final states approach the maximally mixed states. For these areas, there are some points where we cannot obtain their fidelities, because all the populations $p_i^u$ calculated by the Eq. \ref{eq:mj} are zeros. Consequently, we cannot normalize them. The shapes of the contours in \ref{f4}(a) and (c) are similar to those in  \ref{f4}(b) and (d), which confirms the correlations between the efficacy of $\rm M_S$ and the KL divergence $D_{KL}$.
The similarity between the two upper figures and lower figures also demonstrates that, the superiority of correction by $\rm M_S$ is also applicable with the amplitude damping and phase damping noise model.

In conclusion, we have proposed a QEM method that leverages junk information without the need for additional circuits or measurements. In the numerical simulations, we have applied our method to circuits with a few to hundreds of layers. Our method demonstrates significant reductions in infidelities compared to the post-selection method across a wide range of the circuit depth $N$. 
This work has focused on modelling the noise as depolarising errors to deal with practical noise. Parameters in the noise model are extracted by comparing junk measurement results with the model, and errors in other useful measurement results are filtered out accordingly. The same approach can be generalized to models that can more accurately describe the practical noise than depolarising errors.

\vspace{.4cm}

We thank Diandong Tang, Shoukuan Zhao and Jinfeng Zeng for helpful discussions. This work was supported by the National Natural Science Foundation of China (Grants No. 22303005, 12225507, 12088101, 12322413 and 92365206), NSAF (Grant No. U1930403) and Innovation Program for Quantum Science and Technology (No. 2021ZD0301802).

\bibliography{reference}

\newpage   
\clearpage 
\appendix 
\onecolumngrid

\begin{center}
\textbf{\large Supplemental Materials}
\end{center}

\section{Amplitude damping and phase damping noise channel}\label{app:sec1}

\begin{figure}[hbp]
\centering

\includegraphics[scale=1]{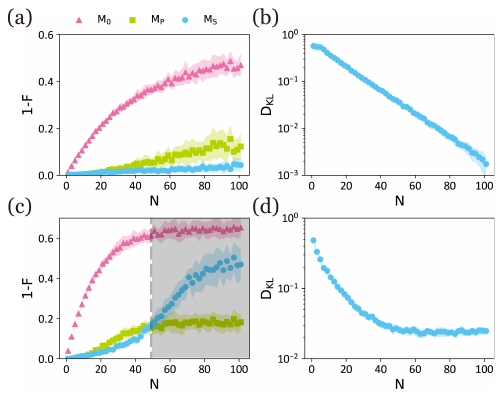}

\caption{\label{supp1} The infidelity and KL divergence $D_{KL}$ with different circuit length $N$ and error rates. The error model consists of amplitude damping noise and phase damping noise on every single- and two-qubit gates without identity gates. The phase damping noise rate on two-qubit gates is set as $\gamma_p = 0.01$. And the amplitude damping rate for two-qubit gates is set as $\gamma_a=10^{-3}$. The corresponding error rate for single-qubit gates is set with $\frac{1}{5}$ of which for two-qubit gates.
The qubit number is $n_q=4$ for (a) and (b), $n_q=8$ for (c) and (d).}
\end{figure}

In this section, we discuss the applicability of $\rm M_S$ with the amplitude damping and phase damping noise channel with different noise strength and qubit numbers.
In Fig. \ref{supp1}(c), the infidelity with three methods is shown when $n_q = 8$. As the qubit number increases, the performance of all three methods worsens, indicating that for larger quantum systems, the gate fidelity should be smaller to operate longer circuits. Additionally, it is observed that before the intersection point, $\rm M_S$ outperforms the other two methods in most of the circuit depth.

\begin{figure}[hbp]
\centering

\includegraphics[scale=1]{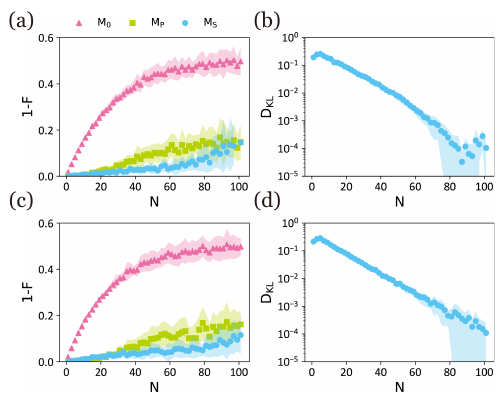}

\caption{\label{supp2} The infidelity and KL divergence $D_{KL}$ are depicted with respect to various circuit lengths $N$ and error rates. The error model incorporates the amplitude damping noise and phase damping noise on every single- and two-qubit gates without identity gates. The qubit number $n_q$ is 4. The phase damping noise rate on two-qubit gates is set as $\gamma_p = 0.01$. And the amplitude damping rate for two-qubit gates is set as $\gamma_a=3\times 10^{-3}$. We assume the single-qubit gates is ideal without errors. For (a) and (b), we do not do the Pauli twirling. For (c) and (d), we add the Pauli twirling to the CZ gates with 20 equivalent circuits for twirling.}
\end{figure}

\begin{figure}[hbp]
\centering

\includegraphics[scale=1]{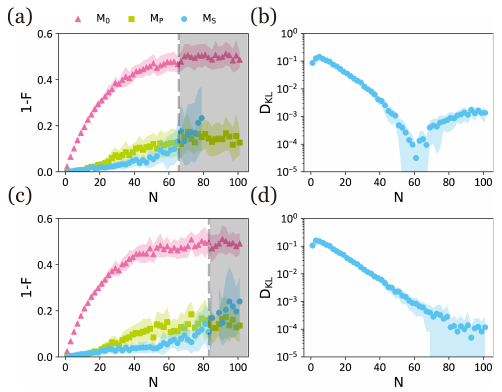}

\caption{\label{supp3} The infidelity and KL divergence $D_{KL}$ are presented across various circuit lengths $N$ and error rates. The error model is the amplitude damping noise and phase damping noise on every single- and two-qubit gates without identity gates. The qubit number $n_q$ is 4. The phase damping noise rate on two-qubit gates is set as $\gamma_p = 0.01$. And the amplitude damping rate for two-qubit gates is set as $\gamma_a=5\times 10^{-3}$. We assume the single-qubit gates is ideal without errors. For (a) and (b), we do not do the Pauli twirling. For (c) and (d), we add the Pauli twirling to the CZ gates with 20 equivalent circuits for twirling.}
\end{figure}

Fig. \ref{supp2} and Fig. \ref{supp3} present the results with larger amplitude damping rates and Pauli twirling. In Fig. \ref{supp2}(a), the amplitude damping rate is 3 times larger than Fig. \ref{supp1}(a). The intersection of $\rm M_S$ and $\rm M_P$ shifts to the left. In Fig. \ref{supp2}(c) and (d), we apply the Pauli twirling technique to change the noise in the circuits to random Pauli noise. We observe a significant improvement for $\rm M_S$.   Increasing the number of random equivalent circuits for twirling may further enhance the performance of $\rm M_S$.

In Fig. \ref{supp3}, the noise rate for the amplitude damping channel increases to 5 times that of Fig. \ref{supp1}. As a result, the performances of the three methods deteriorate. Comparing Fig. \ref{supp3}(a) and (c), we observe a significant rightward shift of the intersection point between $\rm M_S$ and $\rm M_P$ in (c), highlighting the efficacy of the Pauli twirling technique for both $\rm M_S$ and $\rm M_P$ methods. Additionally, for most of the circuit depth, the performance of $\rm M_S$ is significantly better than the other two methods.

\end{document}